# Exploring Children's Use of Self-Made Tangibles in Programming

Alpay Sabuncuoğlu
UNVEST R&D Center
Istanbul, Turkey
asabuncuoglu13@ku.edu.tr

T. Metin Sezgin
KUIS AI Center
Istanbul, Turkey
mtsezgin@ku.edu.tr

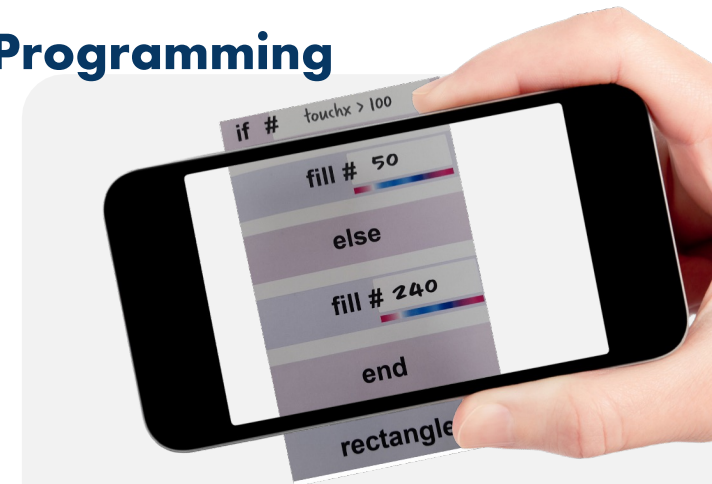

## ABSTRACT

Defining abstract algorithmic structures like functions and variables using self-made tangibles can enhance the usability and affordability of the tangible programming experience by maintaining the input modality and physical interaction throughout the activity and reducing the dependence on electronic devices. However, existing tangible programming environments use digital interfaces to save abstract definitions such as functions and variables, as designing new tangibles is challenging for children due to their limited experience using abstract definitions. We conducted a series of design workshops with children to understand their self-made tangible creation abilities and develop design considerations for tangible computing such as paper programming.

This paper reports:

1. Our insights on how students conceptualize and design tangible programming blocks,
2. Design considerations of self-made tangibles to yield higher understandability and memorability,
3. Tests of our design considerations in creating self-made tangibles in real-life coding activities.

### CSS Concepts

• Social and professional topics → K-12 education; • Applied computing → Education;

### Acknowledgments

The authors gratefully acknowledge that this work was supported by TUBITAK [Grant Number 218K436] and Koç University-İs Bank AI Center. We would like to thank all participant children and their parents for giving their time and effort to made this research possible.

### Notes for Practitioners

*All workshop materials are open-source and can be found at* karton.ku.edu.tr/workshops.

The website includes the presentation and worksheet documents to easily reproduce the activities mentioned in this paper.

We promote educators to share their classrooms' self-made tangible creations with the researchers.

**Our research begins with this question:**

How would a child save the above function via a smartphone camera by building tangibles using craft materials or everyday objects?

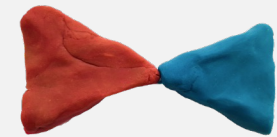

*Using play dough with red and blue colors?*

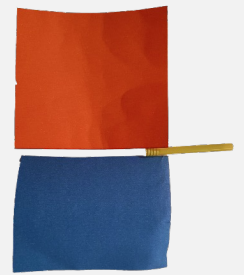

*Using two cardboard pieces with a straw as a separator?*

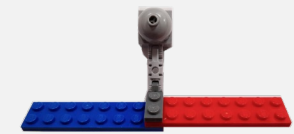

*Using LEGO with a separator that symbolizes the conditional statement?*

Although agreeing upon a precise answer of this question is subjective, at the end of our research, we curated design considerations to help children create well-designed code elements.

## INTRODUCTION

There is a push to integrate programming into the primary education curriculum, given that computational literacy has become a crucial 21st-century ability [4]. This integration aims pupils to acquire vital skills such as creative and computational thinking as well as hands-on skills like building electronics, animations, and integrating technology into daily life [13]. Currently, visual programming environments like Scratch and Alice have become standard tools to teach programming [5, 6]. Interest in tangible programming environments where students code with physical blocks is also growing since, they support inclusivity and collaboration in the classroom environment [9, 11, 12]. However, current tangible environments either require an external coding device like a computer or tablet or costly internal electronics, rendering them out of reach for low socioeconomic levels.

Paper programming environments such as Kart-ON [10], and Budgie [18], bring tangibles to the classroom while making them affordable. In this scenario, students use pre-defined, easily accessible programming blocks to create their programs and run computer-vision-powered devices to scan these blocks. On top of this recently formulated paper programming research, we wanted to explore a programming setup where children create their self-made tangible programming blocks to represent abstract definitions. Research posits that learning occurs when children build and design personally relatable elements [15]. In our scenario, students use everyday objects to associate functions, entities, and properties such as variable assignments and procedure calls. This setup can allow groups of students to create and combine self-made tangible programming blocks for coding, then "run" and observe the behavior of their program using a shared smartphone acting as an interpreter.

We assert that students may actively explore improved ways of framing their algorithms and consider the relationship between commands, inputs, and outputs via self-made tangibles while collaborating with their friends in this screen-free environment. However, integrating self-made tangibles to define abstractions in these paper programming environments brings two main questions:

(1) Is representing abstract programming definitions by self-made tangibles educationally and pedagogically appropriate?

(2) Can self-made tangibles become well-designed code elements that can be memorized and understood by both makers and users of these tangibles in later programming sessions?

We designed a four-step workshop to address these questions where students learned how to define algorithmic tasks, create self-made tangibles, and use them to create programs. In the first step, we teamed up with 7-11-year-old children and decomposed a problem/task together. Next, they designed a set of tangible objects for programming the selected task. In the third step, we tested the memorability and understandability of their self-made tangibles. Based on the test results, we developed design considerations. Finally, we utilized these considerations in a real-life paper programming environment.

This paper presents our quest to integrate student-made tangibles into programming environments. We report our observations, tangible outcomes, findings, and design considerations. We also share our preliminary test results of using design considerations in a real-life paper programming task and discuss the potentials and drawbacks of self-made tangibles for programming.

## SELECTION AND PARTICIPATION OF CHILDREN

We obtained ethical approval for this study from our university's Committee of Human Research. Due to COVID-19 restrictions, we conducted the studies outside with the children from the same neighborhood. We contacted the parents via a large WhatsApp group where parents with primary-school-age kids communicate. We also informed the study details via this channel. This group lives in the same neighborhood in Sarıyer, Istanbul, which has access to technology and quality education. We submitted the informed consent forms via Google Forms for those who responded positively. At the beginning of each workshop, we introduced ourselves, explained the research in a simplified way, and clarified that they were free to go back to their home if they desired. None of the children quit the session or showed stressed behavior.

## RELATED WORK

Between unplugged activities and computationally enhanced tangible programming tools, paper programming offers an affordable way to complete programming activities. Paper programming environments use card-based tangible programming blocks to create a code. Mostly, a separate scanner or a mobile device is utilized to interpret the recognized programming blocks [7, 10]. These environments show the advantages of physical computing while using inexpensive materials as programming blocks. Further, paper programming can be a beneficial step in programming education as it bridges physical computing to scripting languages. This learning approach is coined in "concreteness fading," which discusses an intermediate step between concrete and abstract definitions to help learners distill the generic knowledge and develop conceptual understanding [2].

Existing paper programming environments present a high-level tangible command-set that can be recognized using computer vision algorithms [3, 10]. Although research on using these tools in primary and middle school programming education report several benefits, such as increasing engagement and collaboration, teachers do not adopt them into their curriculum [22]. One limitation of these paper-based tangible programming interfaces is their limited representation ability of tangible input and output variations by not supporting custom tangible blocks to define abstractions. For example, while creating a new program using these tools, users can create a function using the provided tangible materials. However, when they switch to a digital interface to introduce these tangibles as the components of a new function, it can cause an interruption and reduce engagement [3, 10]. But, using self-made tangibles to define abstract definitions in programming languages can enhance the user experience and extend the language expressivity by keeping the same modality and maintaining physical interaction. In order to explore this possibility, we first need to explore students' needs and behaviors with self-made tangibles to integrate this approach in programming environment design effectively.

Tangible User Interfaces (TUI) allow hands-on interaction and control over digital features with physical artifacts [21]. Tangible programming tools can be categorized into three groups from a technical perspective: Electronic, unplugged, and digitally augmented paper-programming kits.

Marshall defines three main advantages of using tangibles in programming [14, 20]:
1. It keeps children mentally and physically active,
2. Enhances collaboration by supporting natural group interactions and increases the visibility of outputs,
3. Naturally invites students to action and improves inclusivity by lowering the threshold for children's participation.

Druin defines four roles that children can partake in a technology design process: user, tester, informant, and design partner [1]. In our workshops, students assumed varying roles, and acted as informants, testers and design partners.

In the informant role, children play a part in the design process at various stages, based on when researchers believe children can inform the design process. As a tester, children are again observed with the technology and asked for their direct comments concerning their experiences. With the role of design partner, children are considered equal stakeholders in the design of new technologies throughout the entire experience.

We started with discovering the interest of children and understanding how they decompose the given problem. Then, we created and tested tangible objects to represent the commands obtained from decomposed problems. Throughout the process, we approach them as design partners.

## TANGIBLE INTERFACE EXAMPLES

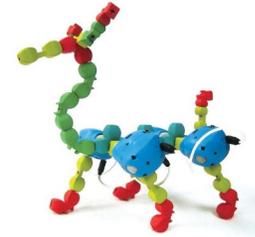

*Topobo uses electronics with motor memory. Raffle et al. 2004. [8]*

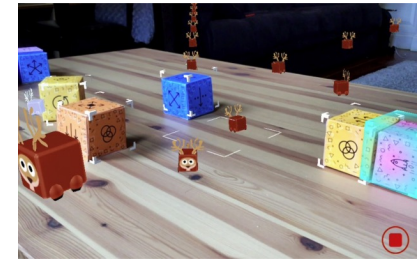

*HyperCubes uses AR to program some motion for 3D elements.. Lleixà. 2018. [3]*

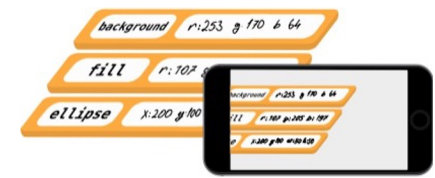

*KartON uses text recognition on smartphone camera. Sabuncuoglu et al. 2019. [10]*

## FLOW OF THE DESIGN STUDIES

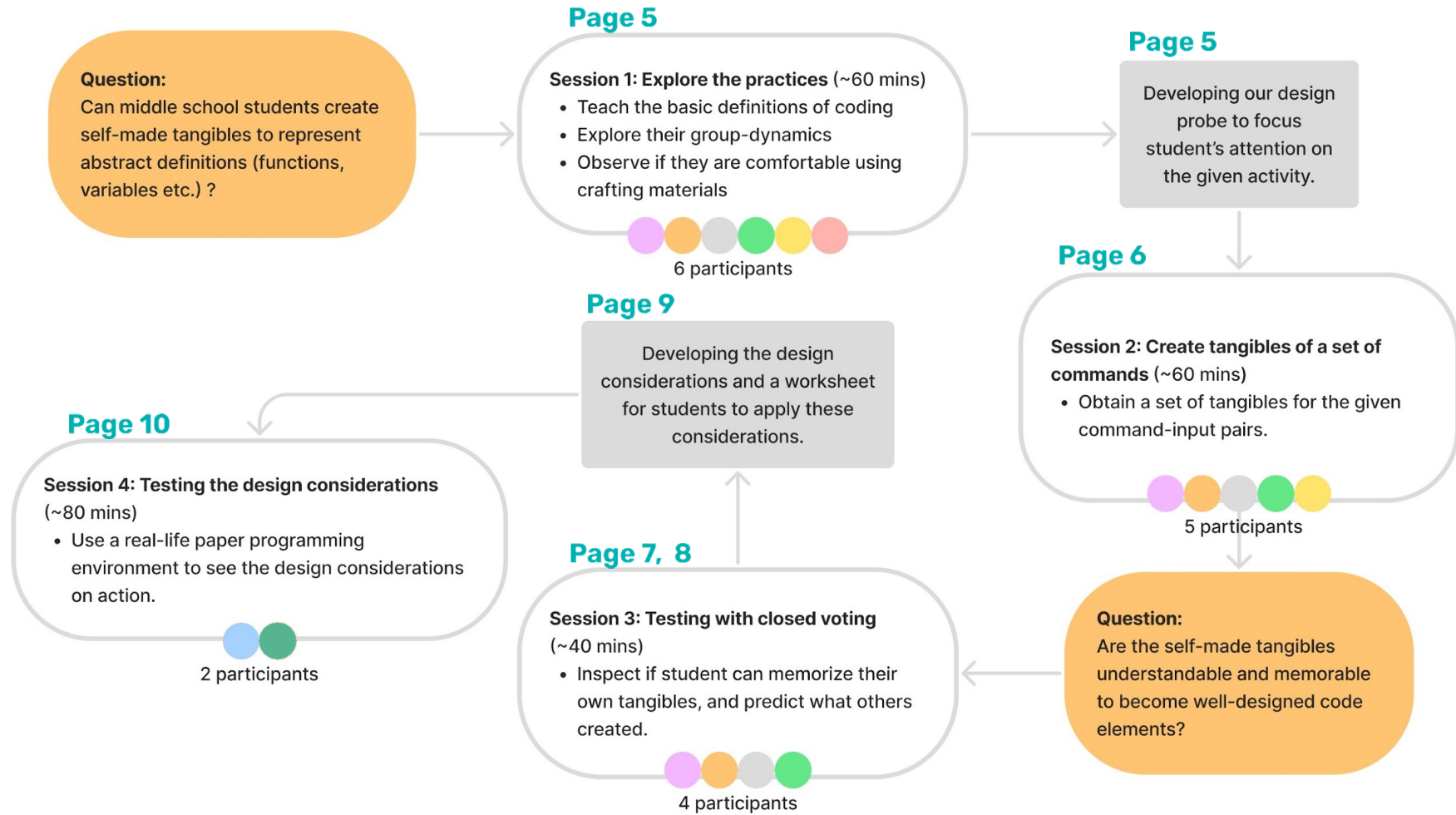

**Limitations of the Studies:** All of the workshops were conducted outdoors due to COVID-19 regulations. Since children have limited experience in programming, we included an introduction before the design process to give the necessary foundation. In total, we had six children of ages 7-11 years old. In Session 1, 2 and 3, we studied with same children, but only four of them remained in all studies. In Session 4, we asked to same Whatsapp group for the participants that did not attend our previous studies, and only two of them participated. Considering participant number and diversity, we cannot easily generalize the findings.

## SESSION 1: INTRODUCING THE CONCEPTS AND LEARNING THE NEEDS

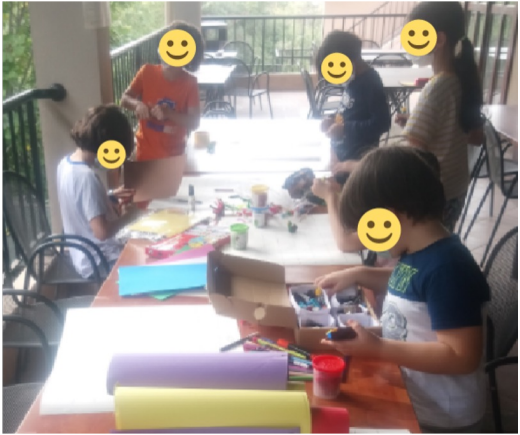

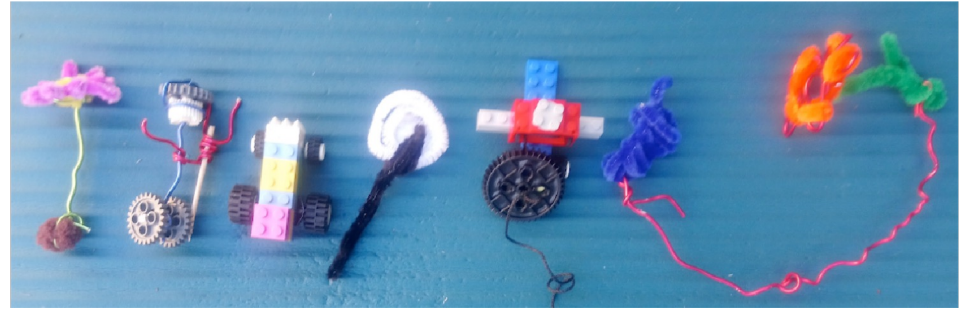

*In the workshop, the provided material box consisted of various items, including paper, clay, wires, pipe cleaners, LEGO pieces, etc. Using these materials, each child created one tangible object per sub-task. Their collaborative output can be seen in this figure.*

We held the first workshop with six children from our local area (one seven-year-old, three ten-year-old, two eleven-year-old children). First, we introduced programming and discussed its role in many breakthroughs in society. Later, we asked them to define a problem or a task to solve with programming. The children came up with the task of "gardening" and started decomposing the possible sub-tasks that could be solved with programming. After listing all the tasks, we agreed upon eight sub-tasks: potting flowers, planting trees, protecting the garden, watering the garden, mowing grass, removing weeds, walking around, and picking fruits.

In this workshop, we wanted to see if children were comfortable with the given materials and their ability to understand the task at hand. Our observations revealed that all students were comfortable with the provided materials and were eager to explore programming concepts regardless of their age. As seen in the above figure, almost all students created tangible objects with different materials. The sizes of the tangibles were similar in each child's creation. Overall, this showed us that our materials were diverse enough to accommodate children's preferences.

**Preparing for the next session:**

Throughout the workshop, we observed some tendency to disengage from the task due to the informality of the workshop. To explain, some children knew each other and also the vicinity. While the familiarity led them to collaborate easily on the given task, it quickly turned into a free chat and distracted them from the task. It is known that group dynamics directly impact children's creativity [19]. Based on this experience, we decided to hand out a design probe for the next workshop to help focus children's activities.

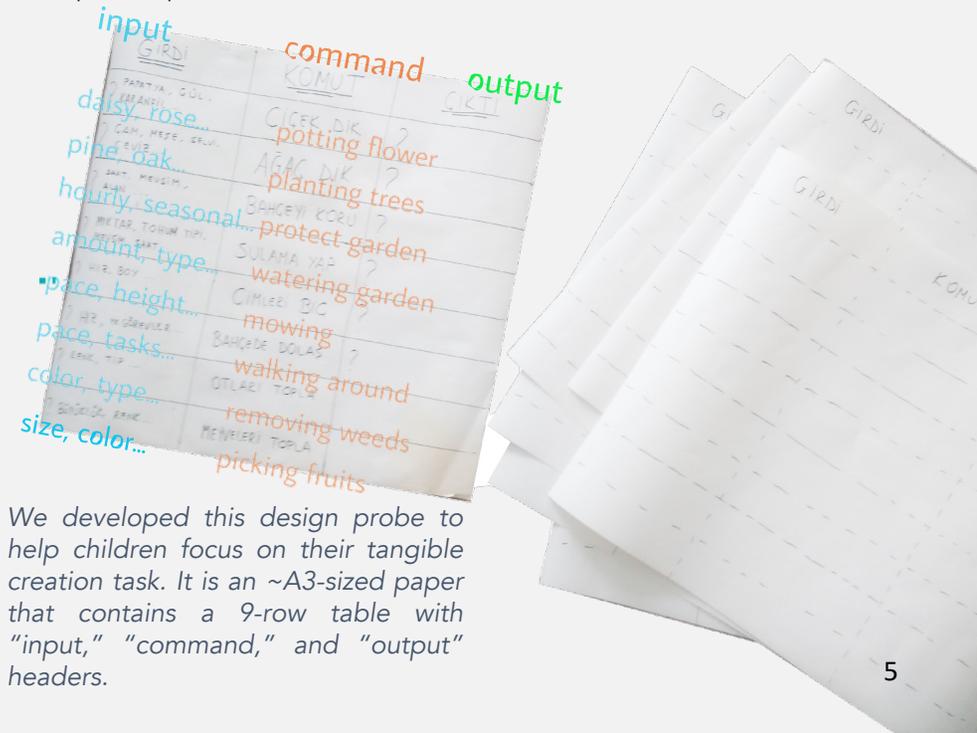

*We developed this design probe to help children focus on their tangible creation task. It is an ~A3-sized paper that contains a 9-row table with "input," "command," and "output" headers.*

## SESSION 2: DESIGNING TANGIBLES OF A SET OF REAL-LIFE COMMANDS

We held the second workshop with five children from the previous study (one seven-year-old, two ten-year-old, two eleven-year-old); one child could not attend the workshop. In this session, we distributed the design probes to help them maintain their focus. We hung a larger copy with commands and possible input names to a wall that all students can see. We handed out empty copies of this probe and asked them to place their final self-made tangibles on pre-defined positions. As intended, the probe helped them to use their time and communicate their ideas more efficiently.

In this session, students created a programming interface for gardening with a total of eight commands, as decided upon earlier. We asked children to design these tangible representations so that a computer (a simple electronic machine) could clearly understand them. But they were not informed about the understandability/memorability test of Session 3. Overall, we obtained forty different programming objects from five children. Here, we shared two distinct samples of these probes that show different input modalities.

After collecting all the objects and seeing how children created these tangibles, we asked if these objects were qualified to become coding elements. A well-designed code element should be easily understood by others and easily memorable by the author of the program. To this end, we tested the self- and inter- memorability and understandability of the programming objects. In this context, we defined understandability as the ability to correctly interpret the meaning of the given tangible objects by other students. And memorability is the ability to remember the represented role of their own tangible object when they see the tangible object later.

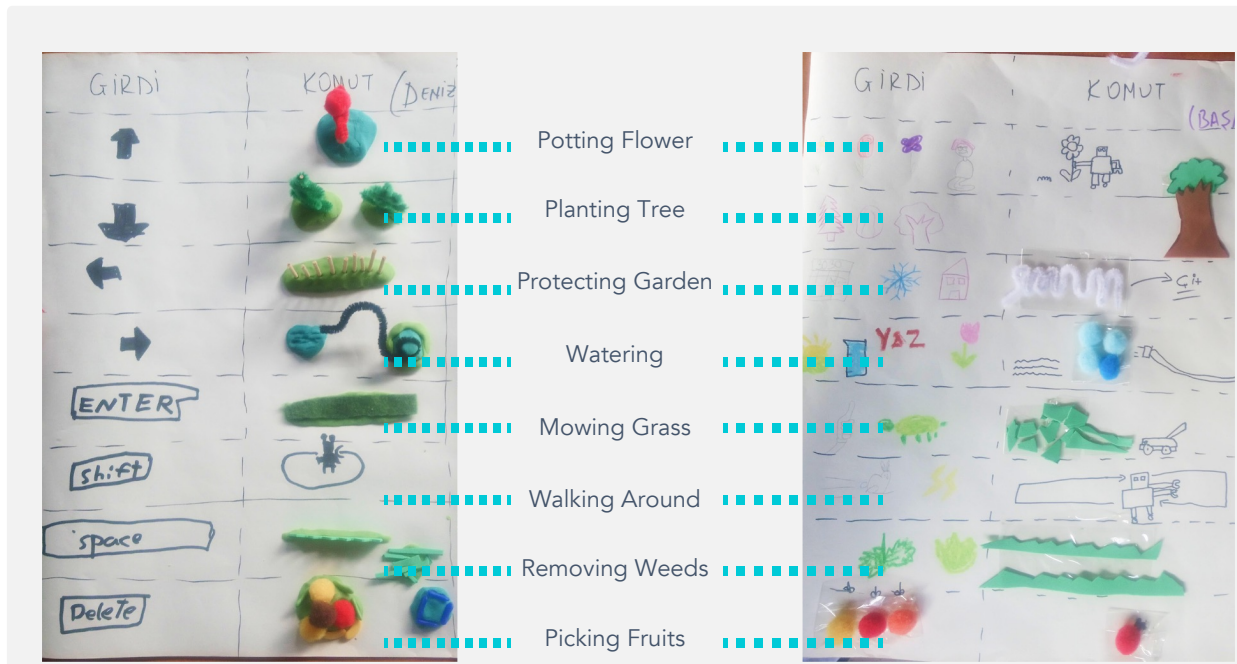

*Deniz offered using keyboard presses to change the given input. He shows the command first, then selects the possible inputs using keyboard shortcuts. Deniz's friend also shared the same design decisions, and resulted in very similar command-input pairs as expected.*

*Başak used a combination of drawing and given materials to create programming objects. Providing input to these commands can take different forms. For example, changing the flower's color in the robot's hand is the input of the "potting flower" command, but watering the garden requires combining two separate drawings.*

## TESTING THE MEMORABILITY AND UNDERSTANDABILITY

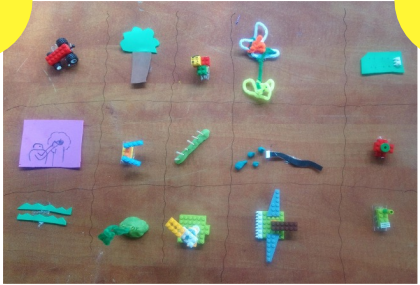
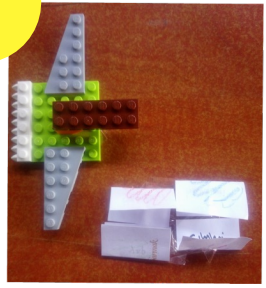

We conducted the test with four children from the previous studies (two ten-year-old and two eleven-year-old). Before children arrived at our workshop area, we prepared a grid of programming objects on a table. We handed color-coded function names to children to test the understandability and memorability of the workshop's outcome.

1. We selected sixteen blocks (four tangible blocks from each participant) and placed them on a table.
2. We distributed color-coded small paper pieces and handed them to the children. Each child is assigned to one color to keep track of recognizing their tangible objects, and they place the enfolded paper pieces onto the grids. They placed this paper onto the grids in a closed form.
3. We discussed whether they remembered these commands or predicted their friends' tangible outputs. We tried to understand the effect of using different materials and different representation methods.

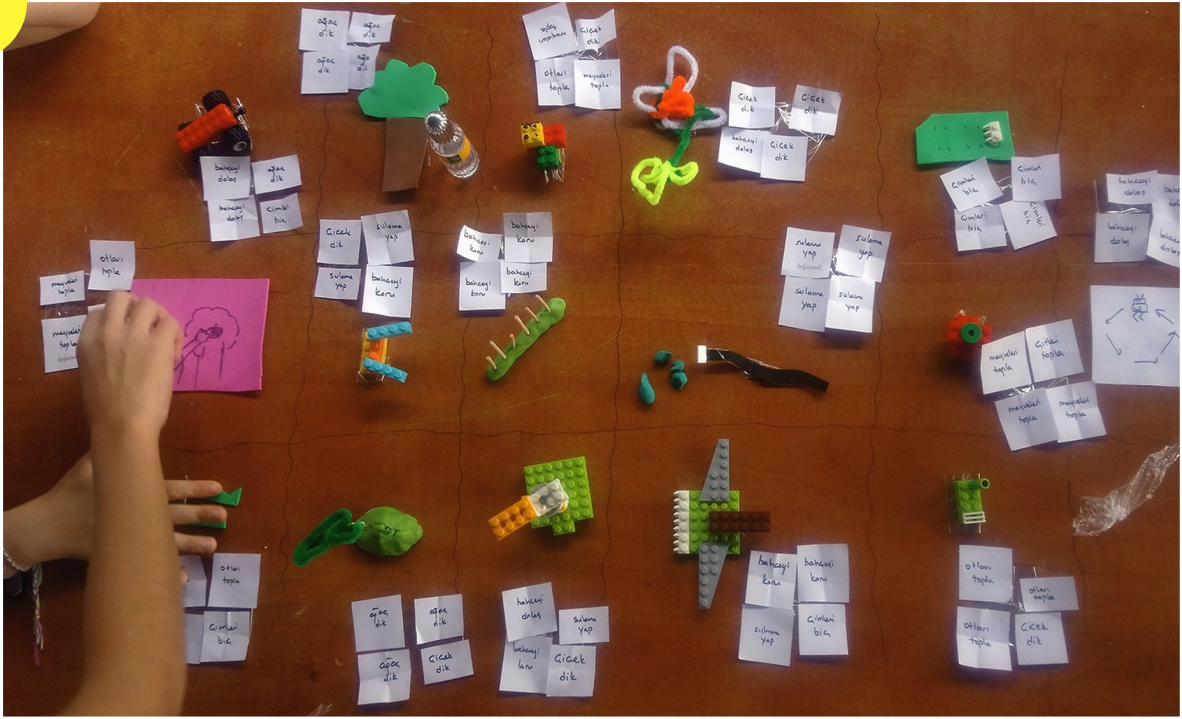

## TEST RESULTS

On the next page, we summarized the test results using a table, which shows the understandability accuracy (top-right corner of each grid), the correct label of the tangible object (bottom-left corner of each square), and each child's answer with color-code information (bottom-right corner of each square).

These results indicate that three main patterns affect the understandability and memorability of a programming object in general:

- Using materials that drive children to create abstract shapes (construction bricks) reduces future understandability.
- The object's recognizability also increases when children depict the action stated in the command rather than directly imitating the object's appearance. Trying to tell the action increases the level of details.
- Similarly, using more than one material increases the level of details.

**Mowing Grass:** Both tangible objects's motivation was using the toothed lego plate as the mower's knives. The top tangible's wings reduced the understandability score.

**Watering:** Representing action as a tangible object with highlighting the correct details generally increases understandability. The blue bricks in the object representation was planned as water cannons, but they were interpreted as protecting gadgets or flowers.

**Picking Fruits:** The LEGO brick representation states nothing (verified by the student; it was random). Surprisingly, this random object is correctly understood by most students.

**Protecting garden:** Both tangibles represent an object related to protecting. The top one is a surveillance camera, and the bottom one is a fence.

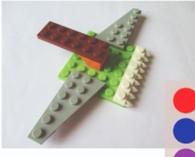

**Potting Flower:** The accuracy of the object at the top was 75% since this student was out of paper pieces, so it was potting flower for this student; however, he couldn't place it.

**Walking Around:** Representing this command with a car seems reasonable, but an abstract car-shaped object can be used for other purposes like mowing and placing seeds.

**Removing Weeds:** Both tangibles represent the object of the action. In this example, mowing and removing weeds have a very similar object. In this case, it is better to represent the action.

**Planting Trees:** Contrast to the previous case, a tree is the only object that is used by commands. So, representing a tree is understandable at first sight.

## DESIGN CONSIDERATIONS

We listed four main design considerations (DC) based on our observation notes, material analysis of student's tangible creations, and memorability/understandability test results. These considerations aim to enrich semantic layers in the building phase of tangible objects to improve the object's understandability and memorability. We also considered the fact that these models will be used in the computer vision systems in real-life programming applications.

**1** Give hints about the action:

The most understandable and memorable designs involved representing the object's visual resemblance and stated action together. For example, "watering a garden" can just be represented using a "hose." But, in our test, we observed that students can interpret this as a protection object and say the command is "protecting the garden." So, adding contextual layers and giving hints about the stated action supports the understandability of the self-made tangible.

**2** Choose the right materials for the environment:

Considering the materials for different locations was an important aspect of the quality of programming time. If the students plan to use the objects more than one-time, using play-dough is not encouraged since it is cracking after drying. Also, considering the fact that these materials will be recognized by a computer vision system in the future studies, light conditions can affect the vision model's accuracy significantly. Similarly, using only wires to represent programming objects result in a high dependency on the foreground- background clarity. Therefore, the environment should be considered while choosing the materials.

**3** Add details with 'simple' shapes:

Adding details helped students to memorize the function of designed tangible. Yet, we emphasized adding details with 'simple shapes' to use students' time more efficiently. In the workshops, they tended to add details with 'fancy' materials and zigzagged scissor moves which distract them from the tasks.

**4** Combine materials:

Combination of materials enhances the interpretability of programming objects from the student perspective. Using combinations of different textures and colors can also increases the accuracy of the computer vision model's recognition.

## MATERIAL ANALYSIS

**Using Play Dough, Wires and Other Shape-Changing Materials:** They are great for adding details, but not long lasting, which decreases the repetitive use of self-made tangibles as an algorithmic structure.

**Using LEGO as Building Blocks:** Five students created a total of 40 tangible function representations, and 15 of these tangibles were completely created with LEGO's. Although students enjoyed using these, the understandability and memorability results are considerably low, compared to other materials.

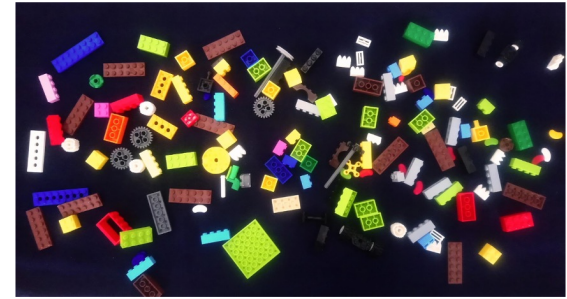

*This photo shows all the LEGO bricks used in the studies. Most of the creations consist of ~4-7 bricks. They combine different colors.*

*In the end, we prepared this worksheet to help children follow some checklists while creating their tangible representations.*

**1** COMMAND — INPUT — OUTPUT

**2** What does happen when you run this command?

**3** Can you create a physical representation of this command using basic shapes?
- Square, rectangle, triangle, circle and ellipse are basic shapes.

**4** Can you use different materials together?
- You can choose LEGO, play dough, wires, pen, cardboard and other materials.

**5** Does your tangible output represent the command's action?

## SESSION 4: APPLYING THE DESIGN CONSIDERATIONS ON PAPER PROGRAMMING CODES

In the previous sessions, children created tangible representations of functions they were already familiar with in daily life, such as 'watering the garden.' Building tangible representations of these functions can be seen as an easier task when compared to more abstract programming blocks like conditional statements or loop structures. In this regard, our goal in the final study was to apply our design considerations to creating tangibles of paper-programming functions and testing their efficacy. First, we gave a programming task and asked children to complete it using Kart-ON programming commands. After creating each code, they ran it to see the output. Then, we asked them to create a tangible representation to save this code as a function.

Two children (Student A and B, both eleven years old) participated in this study. As they did not partake in the previous studies, they were not familiar with the "self-made" programming tangibles and had not heard about our design considerations before. We provided the same materials as preivous sessions (e.g., playdough, LEGO bricks…) and gave 3-5 minutes to create each tangible representation. During the session, we talked with the children about their intended design and helped them to keep track of the worksheet.

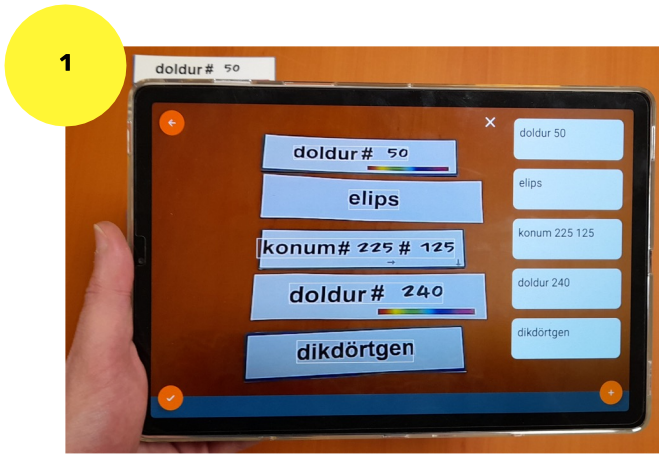

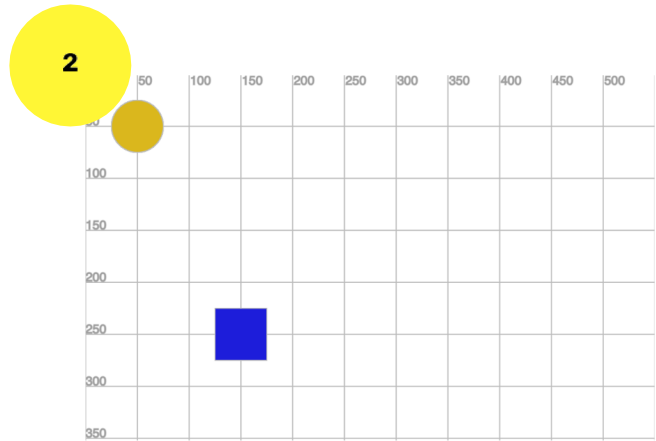

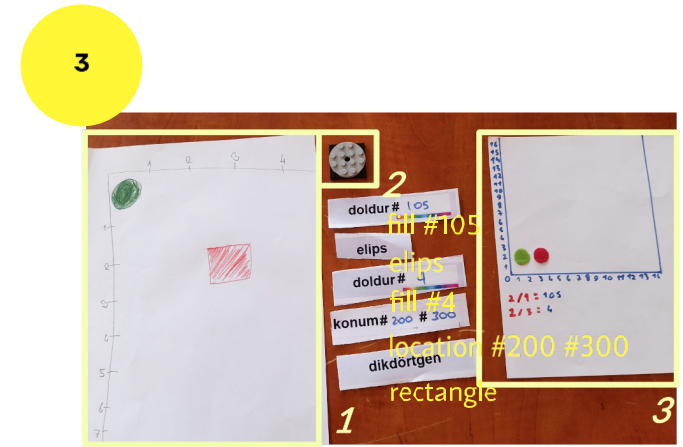

Students used the Kart-ON application, a paper programming environment, to create different algorithmic drawings in this study. This figure demonstrates how the tablet camera captures the paper programming commands When students run the code, they see the output on the left figure. The code (in Turkish) executes the following operations:

- doldur#50: Fills the shape with the color given in Hue value, Hue=50 is amber
- elips: Draws an ellipse with the previously stated color attribute
- konum#225#125: Sets the location of the next shape
- doldur#240: Fills the shape with the color blue
- dikdörtgen: Draws a rectangle with the previously determined attributes

When students tap on the "Compile and Run" button on the bottom-left corner of the screen, they see the drawing above. The application displays the resulted drawing on a coordinate system to help students better understand setting up the locations of the shapes. This figure shows an amber ellipse and a blue rectangle. The ellipse is displayed in its default location, as the user only determined it's color but not its location in the code. The rectangle applied the x and y coordinates of the location command, which are 125 and 225 in the code.

In creating a tangible representation task, both participants tried to replicate the output of the code created earlier on the left in the coordinate system. Box#1 and #2 are created by Student A. The LEGO bricks were added when Student A noticed that he did not apply DC#3. Student B forgot the shape used in the earlier code, which was a square and created a circle with playdough instead (Box#3).

# EARLY INSIGHTS INTO CHILDREN'S TANGIBLE REPRESENTATIONS OF CODES USING DESIGN CONSIDERATIONS

We completed four different creative coding examples in Session 4 (figures below). Here, we summarize early insights of our design considerations in action to create tangible representations for paper programming codes. Although the limited participant number restricts the generalization of our findings, our early observations demonstrated that the playful nature of the tangible building phase made it difficult for students to follow the design considerations.

DC#1 (Giving hints about the action): As Program #1 and #2 do not state any action and only produce static images, students naturally could not give hints about the action. Program #3 was an if-then-else structure requiring checking a touch input, but students did not hint about conditional structure in Box#6 and #7. Both students represented the repeating action for Program#4; Box#8 has rotation capability, and Box#9 has guided students to use materials that can increase lines that demonstrate loop count.

DC#2 (Choosing the right materials): Eventhough we guided students in choosing the durable materials, they tended to use their favorite and more familiar materials, which were LEGO and play dough. Choosing the right materials for future use cases was the hardest design consideration to follow.

DC#3 (Adding details with simple shapes): All tangibles use simple shapes, which helped students to complete the tangible creation task in the given time. We could run the paper codes and create a tangible representation for four different tasks in under eighty minutes.

DC#4 (Combining materials): The first program uses a coordinate system graph and play-dough, which combines different materials and gives details to the users. But, in the second program, Student A did not add the coordinate system and continued with an abstract representation of a snowman. When we reminded him of design considerations, he stated that this version was enough to remember the code. Through the study, the number of combined materials decreased, and students tended to focus on one material in their design.

Overall, these design considerations helped us to manage the duration of the design session. But, students could not fully comprehend nor integrate the design considerations into their representations. We conjecture that they might need more practice and guidance on how to follow them through, which we will consider in the further studies. The challenge of maintaining the balance between playfulness and developing a functional self-made tangible will be addressed in future work.

4. Program #2: Drawing a turquois snowman using the basic circle drawings.

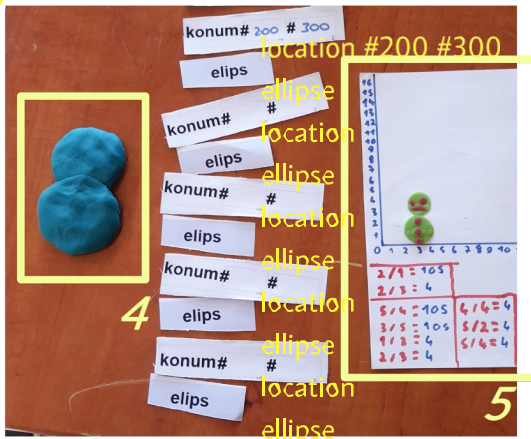

Box#4 has two blue play dough ellipses made by Student A. Yet the student did not want to continue with the coordinate system graph. Student B used the same graph and created a green snowman in Box#5.

5. Program #3: Drawing a triangle either green or red based on touch input.

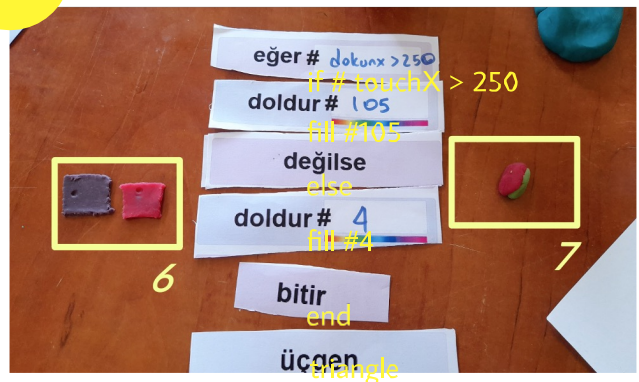

Box#6 uses two play-dough rectangles to show the color changing operation. Box#7 is a one piece play dough that contains two adjacent triangles.

6. Program #4: Drawing 50 consecutive rectangles in a vertical line.

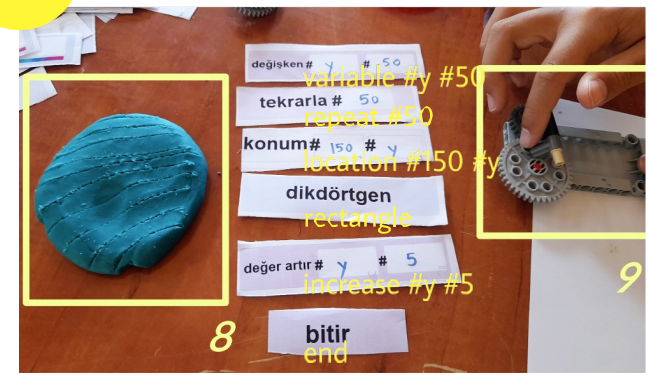

Box#8 have 50 lines that indicates the loop count. Box#9 is a LEGO gear that completes many cycles and stops when the loop count is completed.

## DISCUSSION

Activities that encourage children to design self-made tangibles of abstract definitions can be useful in engaging students and promoting active learning. Throughout the studies, our observations demonstrated that asking students to create tangible representations helps them understand the code, which resonates with the research in active learning [2]. For example, using student-made tangibles in math learning for young children can help them engage in activity, familiarize the concepts in real-world situations and help them understand relational information better. Additionally, using self-made tangibles helped students focus on a code scope's overall working mechanism.

On the other hand, students and teachers need guidance for tangibles to become usable in abstract definitions of programming elements, such as defining new variables and functions. The outputs created in Session 4 indicated that informing children about design considerations may not directly result in improving the quality of tangible objects. Although having a set of considerations helps workshop moderators to guide children more effectively, children can choose to create the tangibles with their favorite materials rather than the logical one. So, rather than delivering these design considerations simultaneously, we suggest practitioners explore these considerations together with children to increase their understandability and memorability in creating tangible representations.

## CONCLUSION

In this paper, we explored the potential use of self-made tangibles in a programming environment. We conducted a total of four sessions that gradually investigated and evaluated tangible programming tasks from decomposing problems to utilizing tangibles with computer vision models. Based on our qualitative observations and analysis, we can summarize the answers of the main research questions as following,

| | |
|---|---|
| *Can students in K3-6 grades build a tangible representation of abstract programming definitions?* | Yes, we observed that students liked building tangible representations of abstract programming definitions, and they could link the function with the representation. |
| *Can students understand and memorize self-made tangible representations in later uses?* | It is not an easy task, even with the design considerations (DC). Our list of DC helped students manage their time and use the materials in a structured way, yet they still had challenges in following them that requires further studies. |

We would like to emphasize that the main aim of these studies was to explore self-made tangibles in designing a language to help students grasp computational thinking skills rather than a complete programming language. Our work is the first step in CCI to explore the creation process and the possible value of self-made tangibles in programming. We conjecture that other researchers may benefit from our accounts and design considerations and extend this line of research further.

## FUTURE WORK

In the final study, we observed that students could follow only some considerations to design more understandable and memorable self-made tangibles. The playfulness of the materials and joy through the creation process overshadowed designing more functional tangibles. One future work path is to help students and educators in following these considerations while keeping the playfulness aspect.

Our next plan is to explore the use of self-made tangibles in more abstract concepts. Throughout the user studies, we used self-made tangibles in physical actions such as coding a high-level robot or an arcade game. For example, "potting a plant" command of a gardening task can be effortlessly translated into an imaginary picture. One might question the possibility of using personally meaningful objects in a more abstract setup. For example, the factorial function is denoted as n! and is equal to $n! = 1*2*3*...*(n-2)*(n-1)*n$, can be recursively defined. As a first step, we can draw the flowchart of this pseudocode to make it more tangible. Then, following the design considerations we created, we can design new or use existing tangible objects to represent the commands. Finally, we will focus on defining a way to link inputs to the commands. For example, we created the structure in the figure below by only using office objects.

Finally, exploring our open research questions can be helpful to extend these workshops' outcomes in an embodied medium, such as Dynamicland.

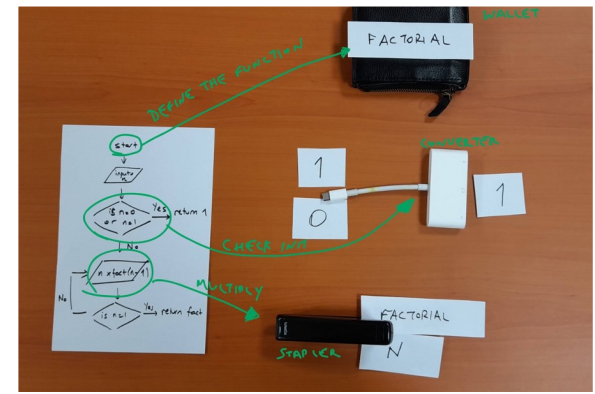